\documentclass[12pt,epsf]{article}

\begin{document}

\begin{flushright}
SLAC-PUB-10072\\
July 21, 2003\\
\end{flushright}

\bigskip\bigskip
\begin{center}
{\bf\large A Cosmological Calculation Suggesting\\a Threshold for
New Physics at 5 Tev\footnote{\baselineskip=12pt Work supported by
Department of Energy contract DE--AC03--76SF00515.}}
\end{center}

\begin{center}
H. Pierre Noyes\\ Stanford Linear Accelerator Center, Stanford
University\\
noyes@slac.stanford.edu\\
\end{center}

\begin{center}
{\bf Abstract}
\end{center}

A calculation by E.D.Jones of the cosmological mass scale for the
phase transition from pre-geometric to physical description as
about 5 Tev could be interpreted as a prediction of an effective
threshold for novel physical effects in particle-particle
collisions.

\begin{center}
\end{center}

\bigskip\bigskip

\begin{center}
To be submitted to {\it Physical Review D} as a brief report
\end{center}

\newpage

E.D.Jones \cite{Jones03,NKLL03} has sketched a compelling
cosmological scenario resting on basic physical principles.
Starting from the fact that the validity of current physics is
bounded, at best, by the Planck length and the Planck density, he
assumes only scaling laws are needed to take the universe from
some pre-geometric, pre-physical situation by an ``extremely
rapid" transition to a much less dense phase in which space, time,
particles and temperature carry their usual meaning. He and we
refer to the end of this transition as {\it thermalization}. The
only parameter which is unknown is the ``number of Plancktons"
--- a {\it Planckton} is a Planck's mass worth of mass-energy at
the Planck density and temperature --- with which, in a poetic
sense, our universe ``starts out". An initial presentation of
Jones' ideas by Noyes, et. al. is available\cite{NKLL03}, with the
{\it caveat} that Jones' own views could differ in ways that have
not yet been spelled out. What makes Jones' work so exciting is
that with this minimal input he is able to show that a currently
acceptable value of $\Omega_{\Lambda}= 0.7$ for the cosmological
constant density normalized to the critical density implies that
the mass scale at which thermalization becomes meaningful is about
5 Tev. In this paper we present a calculation leading to this
result and discuss some of the implications.

That an explicitly gravitational effect has such a low mass scale
is exciting because 5 Tev is well within the energy range that can
be explored once the large hadron collider (LHC) comes on line. In
an alternative approach Thomas\cite{Thomas03,Giddings&Thomas02}
has shown that in 10 dimensional string theories with 6
compactified dimensions and an appropriate choice of the string
coupling constant, black hole effects in hadron-hadron collisions
could become observable at energies as low as 1 Tev, and would
rapidly come to dominate over conventional scattering and particle
production at energies above this threshold. If the threshold is
as low as 4 Tev, effects might also become detectable in very high
energy neutrino-induced, lateral cosmic ray showers\cite{CosRays}
when experimental programs currently under construction collect
enough data. In this paper we argue that Jones' cosmological
calculation might provide us with an estimate of the energy scale
for significant elementary particle gravitational effects in our
universe. Whether this might signal the threshold explored by
Thomas and others, or some even more fundamental reason to
question the adequacy of conventional physics at this mass scale
is briefly discussed below.

Basic operational\cite{Noyes96, Noyes01} criteria for what we mean
by physical measurement limit the highest mass-energy density at
which physical cosmology could possibly be meaningful\cite{NKLL03}
to the Planck density ${3\over 4\pi}M_P^4$. Here $M_P$ is the
Planck mass and we have used units in which $\hbar=1=c=1=k$; in
these units Newton's gravitational constant is given by
$G_N=1/M_P^2$ and the Planck length is $1/M_P$. Because there is
no way to measure geometry, or any type of geometrical structure
at that limit, the only meaningful parameter we can assign to the
universe if it ``starts" at the Planck density is the number of
Plancktons $N_{Pk}$. The same considerations which forced us to
deny the possibility of physical description under these extreme
and necessarily hypothetical conditions require us (following
Jones) to envisage an ``extremely rapid" transition from this
pre-geometrical, pre-physical ``state" to a physically describable
situation. This transition is given explicit form by {\it
defining} the expansion parameter $Z\equiv M_P/\epsilon$ which
takes the universe from the Planck scale $1/M_P$ to a universe
with an event horizon $R_H(\epsilon) =1/\epsilon$. When the scale
factor of the universe reaches this value the
Freedman-Robertson-Walker (FRW) cosmological equations become
appropriate; this fact shows that the parameter $\epsilon$ has
physical significance.

The next step is to note that because no Planckton can be
localized in the pre-geometric situation with which we must start,
each must contribute uniformly to every volume element within the
event horizon. This gives us an alternative way to calculate the
event horizon as the gravitational horizon due to a total mass of
$N_{Pk}\epsilon$, that is a horizon with radial parameter $G_N
N_{Pk}\epsilon ={N_{Pk}\epsilon \over M_P^2}$ . Then we have that
\begin{equation}
1/\epsilon=R_H(\epsilon)={N_{Pk}\epsilon \over M_P^2} \Rightarrow
1={N_{Pk}\epsilon^2\over M_P^2}= {N_{Pk}M_P^2\over M_P^2Z^2}
\Rightarrow N_{Pk}=Z^2
\end{equation}
Thus the number of Plancktons $N_{Pk}$ is not a separate
parameter. It is simply the square of the expansion factor, i.e.
$N_{Pk}=Z^2=M_P^2/\epsilon^2$.

So far all we have done is to envisage some sort of ``expansion"
of the ``virtual energy" which in some sense ``existed" before the
transition was complete to a density where normal space, time and
particles can be defined and provide an appropriate physical
description. But how are we to pick a physical criterion which
will mark the transition from (at best) scaling laws to normal
physics? Jones' answer is that the mass scale at which this
happens is fixed by requiring (momentarily, i.e. at the ``end" of
the phase transition) energy density equilibrium between the
residual energy parameterized by $\epsilon$ and the mass scale,
which we call $m_{\theta}$. The energy $\epsilon$ itself is only
the residual virtual energy {\it per Planckton} ``left behind"
when the bulk of the virtual energy makes a phase transition to
conventional energy at the mass scale $m_{\theta}$ and temperature
$\theta$. Because of the lack of structure at the Planck density,
each of the $N_{Pk}$ Plancktons must contribute $\epsilon$ energy
to each element of volume used in computing the residual energy
density. Hence we require that
\begin{equation}
{3\over 4\pi}m_{\theta}^4 =N_{Pk} {3\over 4\pi}\epsilon^4 \ or \
m_{\theta}^4 =Z^2\epsilon^4
\end{equation}
Of course the thermalized energy emerging from this momentary
equilibrium can be localized down to the mass scale $1/m_{\theta}$
once the residual virtual energy decouples.

We now show that this scenario defines $\epsilon$ as a physically
meaningful parameter. The scenario only makes sense if we can
argue that the virtual energy ``left behind" does, indeed,
decouple from the conventional energy independent of any specific
mechanism used to ``explain" how a pre-physical state expands to a
low enough density so that the phase transition to an FRW state
can take place. But whatever that ``mechanism" is, our postulate
that it represents, in some sense, an expansion from the limiting
(unmeasurable) Planck density to a much lower density where
measurement has at least a conceptual foundation requires that at
the termination of the process it should still be represented
(momentarily) by an ``equation of state" with a ``negative
pressure", i.e. opposing the gravitational self-attraction of the
normal matter. This means (on the grounds of continuity if nothing
else) that, if the transition is ``sufficiently rapid", the
residual energy density ${3\over 4\pi}\epsilon^4$ left behind can
simply be identified with a {\it positive} cosmological constant
density $\Omega_{\Lambda}\rho_c$; here $\rho_c$ is the critical
density which would close the FRW universe in the absence of a
cosmological constant. The positivity of the cosmological constant
follows from the fact that in the FRW universe this sign goes with
negative pressure. ``Sufficiently rapid" amounts to a basic
postulate of the scenario defined by limiting the effects of the
residual energy at the transition point to the identification with
the cosmological constant. Since a positive cosmological constant
in the FRW universe we have now described prevents collapse back
to higher density, the transition is necessarily {\it
irreversible}. As a consequence the Jones scenario makes the mass
parameter $\epsilon$ a physical observable. Explicitly
\begin{equation}
{3\over 4\pi}\epsilon^4 =\Omega_{\Lambda}\rho_c={3\over
4\pi}\left({\Omega_{\Lambda}\over 0.7}\right)\left({h_0\over
0.71}\right )^2\times 5.385\times 10^{-124}M_P^4
\end{equation}
Here we adopt for the value of the critical density\cite{PDG00}
$\rho_c=1.054\times10^{-5}h_0^2 \ Gev/c^2 \ cm^{-3}$. For a
normalized Hubble constant of $h_0=0.71$, we find that $\rho_c =
7.694\times 10^{-124}M_P^4$ in our units. As Eq. 3 indicates, we
take $\Omega_{\Lambda}= 0.7$, a value which is often quoted.
Inserting $\epsilon^4 = M_P^4/Z^4$ into Eq. 3 and solving for Z we
then find that, for $\Omega_{\Lambda}= 0.7$ and $h_0=0.71$,
$Z=6.564\times 10^{30}$. We now solve the energy density
equilibrium equation (Eq. 2) for $m_{\theta}$ and find that
$m_{\theta} =4.766 \ Tev/c^2$.

Note further that, since $M_P/\epsilon = Z = M_P^2/m_{\theta}^2$,
we have that
\begin{equation}
m_{\theta}^2=\epsilon M_P
\end{equation}
or in words, the mass scale for thermalization is the geometric
mean between the cosmological constant ``dark energy" mass and the
Planck mass, {\it independent of Z}! This expression also shows us
that the number of Compton wave lengths of size $1/m_{\theta}$
across the universe of size $1/\epsilon$ is the same as the number
of Planck lengths of size $1/M_p$ across each Compton wave length
$1/m_{\theta}$. But this is the geometrical, counting equivalent
of our argument given above that for uniform density, each
Planckton must contribute mass $\epsilon$ to each volume element
at mass scale $m_{\theta}$, providing consistency with the
intuitive geometrical picture.

To relate these cosmological considerations to elementary particle
physics we now introduce what at first sight will appear to be an
unrelated line of reasoning. Long ago Dyson\cite{Dyson52} pointed
out that if there are $Z_{e^2}=\alpha_{e^2}^{-1}\simeq 137$
electromagnetic interactions within the Compton wavelength of a
single charged particle-antiparticle pair (i.e. $\hbar/2mc$),
there is enough energy to create another pair. Whether these
interactions are virtual, or real (eg in a system with enough
energy and appropriate internal momenta to concentrate
$2mc^2Z_{e^2}$ of that energy within this Compton wavelength), in
a theory for which like charges attract rather than repel each
other still more energy can then be gained by creating another
pair; the system collapses to negatively infinite energy. Dyson
concluded that the renormalized perturbation theory for QED is not
uniformly convergent beyond 137 terms. Note that this bound can be
written as $Z_{e^2}\alpha_{e^2} = 1$. Noyes\cite{Noyes75,Noyes97}
noted that for electron-positron pairs, this critical energy
corresponds approximately to the threshold for producing a pion
because $2m_e\times 137\approx m_{\pi}$. This fact provides a
physical interpretation of the reason for the failure of QED: QED
ignores strong interactions mediated by pions, or more generally
mediated by quarks and anti-quarks which bind to yield pions as
the lowest mass hadronic states.

For gravitation and any mass $m$ the  coupling constant
corresponding to $e^2/\hbar c =\alpha_{e^2}$ is $\alpha_m =G_Nm^2
=m^2/M_P^2$ and the critical condition becomes
\begin{equation}
Z_m\alpha_m = 1 \ or \ Z_m = {M_P^2\over m^2}
\end{equation}
where $Z_m$ represents the number of gravitational interactions
within $\hbar/mc$ defining this critical condition. That is, for
quantum gravitational perturbation theory, the cutoff mass-energy
corresponds to the Planck mass rather than the pion mass, which
makes sense in an elementary particle context. But this means that
at the moment of the phase transition at mass scale $m_{\theta}$
in the cosmological context we have been discussing,
$Z_{m_{\theta}} = M_P^2/m_{\theta}^2 =Z$. That is, at the mass
scale $m_{\theta}$ --- which we can predict, given the
cosmological constant density or equivalently the expansion factor
$Z$ from the Planck density --- this same $Z$ is {\it also} the
Dyson-Noyes gravitational saturation parameter for particles of
mass $m_{\theta}$. Recall now that above we showed that, at this
mass scale, there are just enough systems of this mass and Compton
wavelenght $1/m_{\theta}$ to, geometrically, fit into the event
horizon defined by $\epsilon$. But each such Compton wavelength
itself defines an event horizon within which (barring the
existence of some heavier mass) no structure can be defined using
particle probes. This again establishes the self-consistency of
the scenario provided that {\it at the moment of thermalization}
there are no heavier particulate masses which can be given
physical meaning.

If one accepts HPN's reasoning that explains the Dyson breakdown
of perturbative QED at 137 terms as due to the production of pions
of mass $m_{\pi} \approx 2m_e/\alpha_{e^2}\approx 274 m_e$, we
might by analogy say that perturbative gravitational theory breaks
down at around 5 Tev because it leaves out the production of some
(currently unknown) particle of mass $m_{\theta}$. One obvious
candidate is that this mass is the characteristic mass of
particulate dark matter, which at least to a first approximation
has only gravitational interactions. This was already suggested
earlier\cite{NKLL03}, but not supported by the arguments given
here. If this identification is correct, particulate dark matter
searches might eventually pick it up. Of course, if this
interpretation is to be consistent, such a particle would have to
have a structure which would stabilize it for several Gigayears,
which requires a theory that we do not pursue further here.
Obviously it could have interesting gravitational effects at LHC
energies, but could {\it not} be a black hole; conventional black
holes of this mass would be unstable due to Hawking radiation.
Theorists using string theory explore the possibility of detecting
black hole production at the LHC
(\cite{Thomas03,Giddings&Thomas02} and others there cited) and in
very high energy neutrino-induced lateral cosmic ray showers
(eg\cite{CosRays} and others there cited), might find it
profitable to take account of this cosmological significance of
the 5 Tev energy range in their parameter studies.

In conclusion we note that the three parameters of Jones' theory
of {\it Microcosmology}, namely $\epsilon$, $Z^2=N_{Pk}$ and
$m_{\theta}$, emphasize different ways of looking at the theory
once we accept it as an overall description of our universe. From
an empiricist's point of view, measuring $\epsilon$ determines (at
least approximately) the number of Plancktons with which our
universe starts out and predicts the mass-energy-temperature scale
at which the FRW description first becomes appropriate. Like any
boundary condition, this leaves unanswered the question {\it why}
this parameter describes our universe.

If we take the basic parameter to be $N_{Pk}$, the unanswered
question remains the same, but this point of view raises the
possibility of exploring to what extent the cosmological
consequences of the choice of the single parameter $N_{Pk}$ set
constraints on the physical parameters (eg. elementary particle
coupling constants) by requiring them to be consistent with
cosmological thermalization starting at 5 Tev.

If we focus our attention on $m_{\theta}$, we naturally are led to
ask {\it why} this particular energy is singled out. Many
theoretical speculations are possible. One
suggestion\cite{NKLL03}, already discussed briefly above, is that
$m_{\theta}$ is the mass of particulate dark matter. Then the
saturation of the low energy gravitational interaction, or bound
on range of validity of perturbative quantum gravity, would
correspond to the threshold for production of this particle. Here
we assume the correctness of our interpretation that the Dyson
calculation of the bound on the validity of perturbative QED
corresponds to the threshold for pion production, and infer that
particulate dark matter plays the same role in gravitational
theory. Clearly this also would imply that 5 Tev is a significant
threshold for new physics. An alternative already mentioned is
that what has been cosmologically calculated by Ed Jones is in
fact the threshold for black hole production allowed by a
particular choice of the number of compactified dimensions and
string coupling constant\cite{Thomas03,Giddings&Thomas02}.

Independent of any specific theory, we emphasize the fact that 5
Tev is well within the range which will be opened up to {\it
experimental} study once the large hadron collider (LHC) comes on
line without endorsing any specific theoretical interpretation. We
believe that Jones' cosmological calculation greatly strengthens
the expectation that {\it fundamental new physics} can be expected
to emerge from the work at that installation.

In closing this author wishes to thank, once again, E.D.Jones for
sharing his ideas and permitting us to discuss them prior to
posting his own work. This author is also much indebted to V.A.
Karmanov for extensive help in clarifying the physical ideas
presented here. He also wishes to thank his collaborators in the
earlier discussion (L.H.Kauffman, J.V.Lindesay, and
W.R.Lamb)\cite{NKLL03} for permission to extend our joint work in
this particular way.

\end{document}